\documentclass[aps]{revtex4}
\usepackage [cp1251]{inputenc}
\usepackage{amsmath}
\usepackage{bm}
\usepackage{graphicx}
\usepackage{bm}
\usepackage{epsfig}
\usepackage{epstopdf}
\usepackage{amsmath}
\usepackage{color}

\begin{document}

	\title{Theory of acoustic-phonon involved exciton spin flip \\in perovskite semiconductors}
	
	\author{A.V. Rodina\footnote{anna.rodina@mail.ioffe.ru} and E.L. Ivchenko}
	
	\affiliation{Ioffe Institute, 194021 St.~Petersburg, Russia}

	\begin{abstract}
	We present a theory of the acoustic phonon assisted spin-flip Raman
scattering (SFRS), or resonant
photoluminescence with the spin flip of a photoexcited exciton localized
in a bulk cubic-phase
perovskite semiconductor. We consider the spin-flip transitions between
the ground-state exciton
spin sublevels in external magnetic field ${\bm B}$ and discuss the
variation of their probability rate
and polarization selection rules with the increase of ${\bm B}$. The
transitions are treated as two-quantum
processes with the virtual to and fro transfer of the electron in the
electron-hole pair between the bottom and
first excited conduction bands. The transfer occurs due to both the
electron-hole exchange interaction and
the electron-phonon interaction. The theoretical results allow one to
distinguish the phonon assisted
Raman scattering from (a) the resonant Raman scattering with the
combined spin flip of the localized
resident electron and hole and (b) the biexciton-mediated SFRS analyzed
previously.


	%
	\end{abstract}


	\date{January 29, 2024}
	
	\maketitle

\section{Introduction}

Spin-flip Raman scattering (SFRS) of light via the resonant excitation of excitons is a striking example of the light-matter interaction in bulk and nanoscale semiconductors. The light scattering is accompanied by a reversal of one or a few charge-carrier spins or of the exciton spin as a whole in an external magnetic field. Usually the SFRS spectra show single spin-flip lines, while double or multiple reversals of carrier spins are rarely observed. In nanostructures, the spin flips of two resident electrons have been first observed experimentally and described theoretically in colloidal quasi-two-dimensional CdSe nanoplatelets \cite{Kudlacik2020,Rodina2020}. Since then, resonant light scattering with simultaneous spin flips of the localized electron and hole has been registered in perovskite bulk (3D) crystals \cite{Kirstein2022,Kopteva2023} and nanocrystals (0D) as well as in new type of two-dimensional (2D) lead halide perovskites \cite{Harkort2023}.

In the previous work~\cite{Rodina2022}, we investigated resonant Raman scattering of light in semiconductor perovskites in an external magnetic field. The following  scattering process with spin flip of resident electrons and holes was considered: (i) resonant excitation of  a localized exciton (or exciton polariton) by incident light with frequency $\omega_0$, (ii) the spin flip of a resident charge carrier due to the exchange interaction of the resident electron with an electron in the exciton or the resident hole with a hole in the exciton, and (iii) the exciton emission of the  secondary photon with the energy $\hbar \omega$  shifted from $\hbar \omega_0$ by the Zeeman splitting of the resident quasi-particle $g_e \mu_{\rm B} B$ or $g_h \mu_{\rm B} B$, where $g_e$ and $g_h$ are electron and hole effective $g$ factors. In addition to the single scattering processes $1e$ and $1h$, we have also calculated the efficiency of the double scattering ($1e,1h$), in which the Raman shift $\hbar (\omega_0 - \omega)$ is equal to the sum or difference of the single-particle Zeeman splittings $(\pm g_e \pm g_h) \mu_{\rm B} B$. The paper~\cite{Rodina2022} also analyzes the biexciton mechanism of light scattering. The latter does not require the presence of resident electrons or holes but assumes a sufficiently large light intensity to accumulate in the sample a noticeable nonequilibrium concentration of excitons excited by the same primary radiation. In the biexciton mechanism, the interaction of this radiation with the nonequilibrium excitons leads to the flipping of the exciton spin as a whole. In Ref.~\cite{Rodina2022} we pointed to an additional mechanism of light scattering with an exciton spin-flip: an exciton excited by the incident light emits or absorbs an acoustic phonon, after which the secondary light is emitted at the frequency $\omega = \omega_0 \pm g \mu_{\rm B} B/\hbar$, where $g=g_e + g_h$ is the exciton $g$ factor. In fact, this process can be considered as the resonant Brillouin scattering. Interestingly, although the electron and hole $g$ factors in perovskite semiconductors can vary significantly demonstrating the universal dependence of the band gap value $E_g$ \cite{Kirstein2022}, the value of the exciton $g$ factor $g=g_e + g_h$ is found to be nearly constant, ranging from +2.3 to +2.7 \cite{Kopteva2023}.

In the present work, a theory of the acoustic phonon mechanism of resonant light scattering with exciton spin flipping is developed. The scattering efficiency with the Raman shift proportional not only to the value of $g\mu_{\rm B} B$ but also to its half is calculated. The latter possibility allows one to distinguish this mechanism from the previously considered spin flips of noninteracting resident electron and hole. 

The paper is structured as follows. In Sec. \ref{setup} we introduce the ground exciton states in a cubic perovskite and present the matrix elements of the electron-hole exchange interaction between ${\cal R}^-_6 \times {\cal R}^+_6$ and ${\cal R}^-_8 \times {\cal R}^+_6$ electron-hole pairs.  In Sec. \ref{III} we derive the matrix elements of the deformation potential operator between the electron
states  ${\cal R}^-_6$ and ${\cal R}^-_8$. Section \ref{IV} concerns the compound matrix element of the exciton spin flip mediated by an acoustic phonon and the electron-hole exchange interaction,  scattering-fluorescence duality, and the scattering efficiency. Section \ref{DC} presents the discussion and conclusion.

\section{EXCITON STATES AND ELECTRON-HOLE EXCHANGE INTERACTION IN
CUBIC PEROVSKITES} \label{setup}

The ground state excitons in the perovskite semiconductor with the cubic phase are formed from the electron ${\cal R}_6^-$  states at the bottom of the conduction bands with total momentum $j=1/2$  and the hole  ${\cal R}_6^+$  states  at top of the valence bands with spin $s=1/2$  of the ${\cal R}$-point. The wave functions of the exciton ground state transform according to the representation $D_{\rm exc}^{(6,6)}= {\cal R}_6^- \times {\cal R}_6^+$ which is reducible and may be decomposed into irreducible representations $A_{2u} + T_{1u}$ of the point group O$_h$. The electron-hole exchange interaction partially removes the degeneracy and splits the 1$s$-exciton level into a triplet with the angular momentum $J=1$ (the representation $T_{1u}$) and a singlet (the representation $A_{2u}$, the angular momentum $J=0$), Fig. 1. The higher conduction band has the ${\cal R}_8^-$ symmetry and is characterized by the  total momentum $j=3/2$. The  wave functions of the excited electron-hole pairs transform according to the representation $D^{(8,6)}= {\cal R}_8^- \times {\cal R}_6^+$ which is also reducible and may be decomposed into irreducible representations $E_{u}+ T_{1u} + T_{2u}$ of the point group O$_h$.

We introduce the basis functions for ${\cal R}_6^+$ , ${\cal R}_6^-$ and ${\cal R}_8^-$ bands  as:
\begin{eqnarray}
\label{basise-}
&& \psi^{(e)}_{{\cal R}_6^-, 1/2} = - \frac{1}{\sqrt{3}} [ \alpha Z + \beta (X+ {\rm i} Y) ]\: , \\
&& \:\psi^{(e)}_{{\cal R}_6^-, -1/2} =\frac{1}{\sqrt{3}}\: [ \beta Z -
\alpha (X- {\rm i} Y) ] \:, \nonumber \\ \label{basish+}
&& \psi^{(h)}_{{\cal R}_6^+, 1/2}  =~  \alpha S\;, \quad \psi^{(h)}_{{\cal R}_6^+, -1/2} = \beta S\:,
\end{eqnarray}
and
    \begin{eqnarray} \label{basis8}
&& \psi^{(e)}_{{\cal R}_8^-, 3/2}  = - \alpha  \frac{X+ {\rm i} Y}{\sqrt{2}}\:,  \\ && \psi^{(e)}_{{\cal R}_8^-, 1/2}  = \frac{1}{\sqrt{6}} \left[ 2 \alpha Z - \beta (X+ {\rm i} Y) \right] \nonumber\:,\\ &&\psi^{(e)}_{{\cal R}_8^-, -1/2} = \frac{1}{\sqrt{6}} \left[ 2 \beta Z +  \alpha (X- {\rm i} Y) \right] \:,\nonumber\\ && \psi^{(e)}_{{\cal R}_8^-, -3/2}  = \beta \frac{X- {\rm i} Y}{\sqrt{2}}\: \nonumber\:.
\end{eqnarray}
Here $S$ is an invariant orbital Bloch function, and $X, Y, Z$ are Bloch functions transformed as coordinates $x,y,z$ by operations of the point group O$_h$; $\alpha$ and $\beta$ are two-component spin columns for spin states with 1/2 and $- 1/2$ projection on the $z$ axis being the eigen functions of the Pauli matrix $\sigma_z$. The energy band structure of cubic phase perovskites near the ${\cal R}$ point is shown in Fig. 1(a). The ${\bm k}$-dispersion of the ${\cal R}_6^+$ band describes the holes in the valence band (h), ${\cal R}_6^-$ --  ground state electrons in the conduction band (e) and ${\cal R}_8^-$  -- light (le) and heavy (he) electrons in the higher conduction band.

The triplet and singlet ground states can be presented in the form
\begin{equation}\label{Phi}
\Psi^{(1s)}_{1,j} = \Phi({\bm r}_e,{\bm r}_h) \psi^{(6,6)}_{1,j}~(j=x,y,z)\:,\: \Psi^{(1s)}_{0,0} =  \Phi({\bm r}_e,{\bm r}_h) \psi^{(6,6)}_{0,0}\:,
\end{equation}
where the two-particle envelope  wave function $\Phi(\bm r_e,\bm r_h)$ describes the ground state of the ${\cal R}^-_6 \times {\cal R}^+_6$ exciton localized as a whole on potential fluctuations or at a  defect in a bulk crystal or in the nanostructure. If the exciton localization energy is small compared to the exciton binding energy then the envelope can be factorized
\begin{equation} \label{fact}
\Phi({\bm r}_e,{\bm r}_h) = F({\bm r}) \phi({\bm r}_e - {\bm r}_h)\:,
\end{equation}
where ${\bm r}$ is the radius-vector of the exciton center of mass. Then, for coinciding coordinates of the electron and hole, one has $ \Phi ({\bm r},{\bm r})=\phi(0)F(r)$.
The two-particle basis functions $\psi^{(6,6)}_{1, j}$ are  the products of the electron and hole Bloch functions at the ${\cal R}$-point that transform as coordinates $x,y,z$ (representation $T_{1u}$ of the $O_h$ group) and are given by
\begin{equation}\label{6xyz}
\psi_x^{(6,6)}  = \frac{1}{\sqrt{2}} ( \psi_{1,-1}^{(6,6)}  -  \psi_{1,1}^{(6,6)} ) \:,\:
\psi_y^{(6,6)}  = \frac{\rm i}{\sqrt{2}} ( \psi_{1,-1}^{(6,6)}  +  \psi_{1,1}^{(6,6)} ) \:,\: \psi_z^{(6,6)}  =\psi_{1,0}^{(6,6)}  \:,
\end{equation}
where
\begin{eqnarray} \label{1100}
&&\psi^{(6,6)}_{1,1} = \psi^{(e)}_{{\cal R}_6^-, 1/2} \psi^{(h)}_{{\cal R}_6^+, 1/2}\, ,
\quad \psi_{1,-1}^{(6,6)} = \psi^{(e)}_{{\cal R}_6^-, -1/2} \psi^{(h)}_{{\cal R}_6^+, -1/2}\:, \nonumber  \\ &&\psi_{1,0}^{(6,6)} = \frac{1}{\sqrt{2}}  \left(\psi^{(e)}_{{\cal R}_6^-, 1/2} \psi^{(h)}_{{\cal R}_6^+, - 1/2} + \psi^{(e)}_{{\cal R}_6^-, -1/2} \psi^{(h)}_{{\cal R}_6^+, 1/2} \right) \nonumber\:.
\end{eqnarray}
The states (\ref{6xyz}) are optically active in the dipole approximation, while the $A_{2u}$ state
\begin{equation}
\psi_{0,0}^{(6,6)} = \frac{1}{\sqrt{2}}  \left(\psi^{(e)}_{{\cal R}_6^-, 1/2} \psi^{(h)}_{{\cal R}_6^+, - 1/2} - \psi^{(e)}_{{\cal R}_6^-, -1/2} \psi^{(h)}_{{\cal R}_6^+, 1/2} \right) \nonumber
\end{equation}
is optically forbidden.

The short-range electron-hole exchange interaction is described by the Hamiltonian
\begin{equation} \label{hexch}
{\cal H}_{\rm exch} = \hat h_{\rm exch} \delta({\bm r}_e - {\bm r}_h)\:,\:  \hat h_{\rm exch} = w\hspace{0.5 mm}{\bm \sigma}^e \cdot {\bm \sigma}^h  \:,
\end{equation}
where $w=w_0 \Omega_0$, $w_0$  is an energy  exchange constant and $\Omega_0$ is  the volume of the unit cell, $\sigma_{\alpha}^e$ and $\sigma_{\alpha}^h$ ($\alpha=x,y,z$) are the Pauli matrices acting on the spin columns $\alpha, \beta$ of the electron and hole, respectively. The matrix elements of these operators taken between the basic states (\ref{basise-}) and (\ref{basish+}) have the form
\begin{eqnarray} \label{matrelem}
\langle \psi^{(e)}_{{\cal R}_6^-, s'} |{\bm \sigma}^e | \psi^{(e)}_{{\cal R}_6^-, s} \rangle = - \frac13 {\bm \sigma}_{s's}\:, \quad
\langle  \psi^{(h)}_{{\cal R}_6^+, s'} | {\bm \sigma}^h |  \psi^{(h)}_{{\cal R}_6^+, s} \rangle = {\bm \sigma}_{s's}\:. 
\end{eqnarray}
The exchange interaction \eqref{hexch} causes the splitting
$$\Delta_{10} = 4 w \int d {\bm r} |\Phi({\bm r},{\bm r})|^2 =  4 w \phi^2(0)  $$
between the exciton states $T_{1u}$ with $J=1$ and $A_{2u}$ ($J=0$). Here  the function $\phi ({\bm r}_e - {\bm r}_h) $ describes the relative motion of the electron and hole in the ground state exciton.

\begin{figure*}[h]
\includegraphics[width=0.75\textwidth]{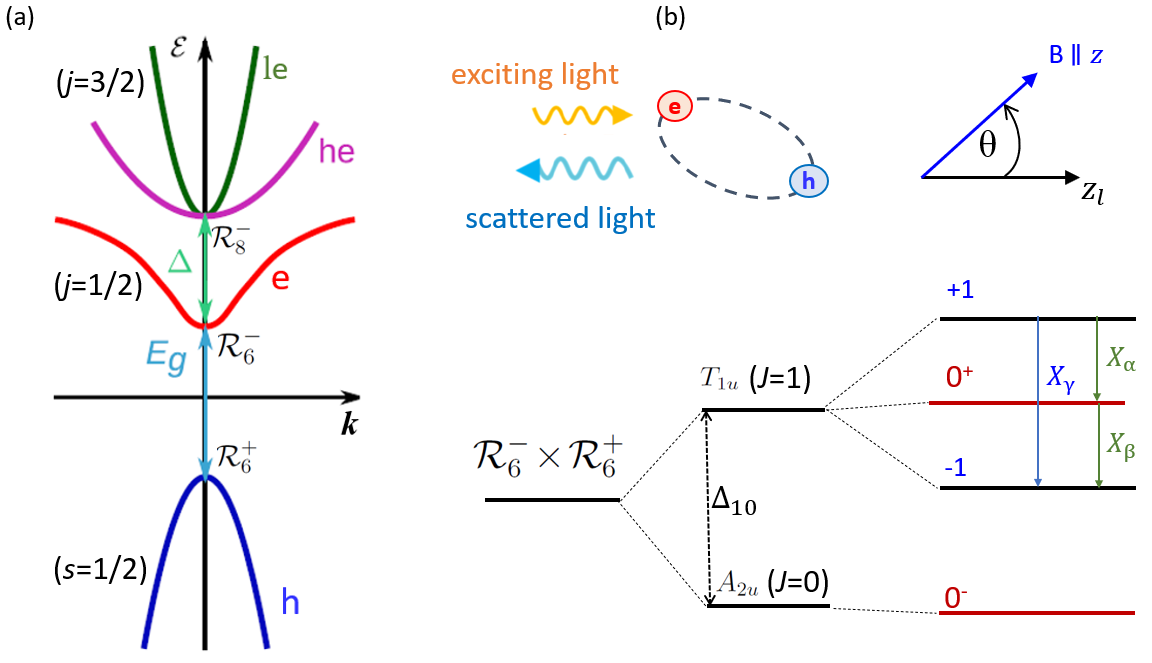}
   \caption{ (a) Schematic band structure of cubic perovskite semiconductors.  (b) Schematic of the exciton ground-state levels at zero and weak external magnetic fields. Vertical arrows show  the light scattering processes with the spin-flip transitions $X_{\alpha}$: $|+1\rangle \equiv (J=1, j_z = 1)$ to $|0^+\rangle  \equiv  (1,0)$; $X_{\beta}$: $|0^+\rangle   \to |-1\rangle  \equiv (1,-1)$ and $X_{\gamma}$: $|+1 \rangle \to |-1\rangle $.}
 \label{Fig1}
\end{figure*}
\begin{figure*}[h]
\includegraphics[width=0.65\textwidth]{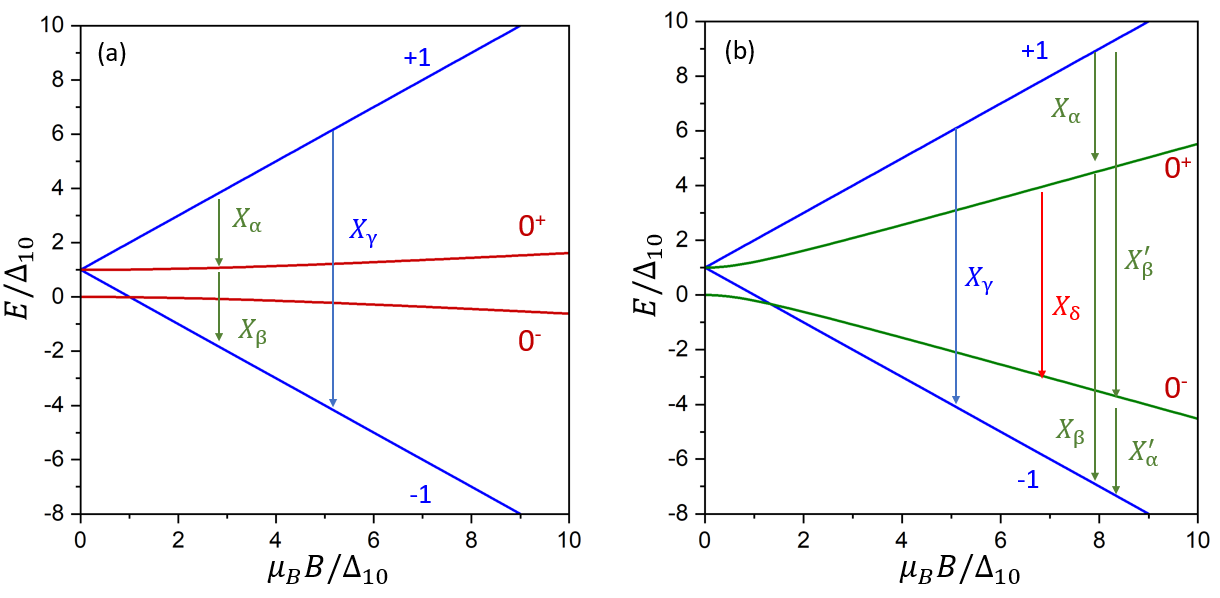}
\includegraphics[width=0.65\textwidth]{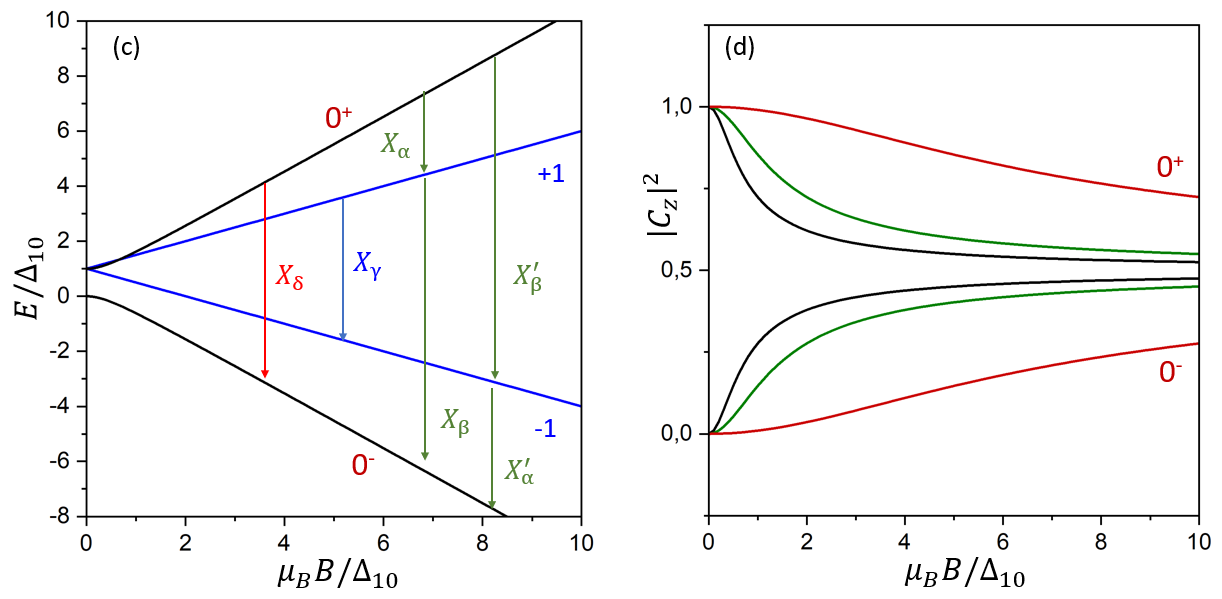}
   \caption{ Evolution of the exciton ground level splitting in external magnetic fields for (a) $g_e+g_h=2$, $g_e - g_h=0.2$, (b) $g_e+g_h=2$, $g_e-g_h=1$, and (c) $g_e+g_h=1$, $g_e-g_h=2$. The blue, green and red vertical arrows show the spin-flip transitions $X_\gamma$ with $\Delta j_z=2$, $X_{\alpha}$, $X_{\beta}$, $X'_{\alpha}$, and  $X'_{\beta}$ with $\Delta j_z=1$, and $X_{\delta}$ with $\Delta j_z=0$, respectively, as described in Section \ref{DC}. Panel (d) shows the variation of the bright exciton $\psi^{(6,6)}_{1,0}$ contribution to the ground exciton states $|0^+ \rangle$ and $|0^-\rangle $ with zero projection on the magnetic field for  $g_e-g_h=0.2$ (red lines), 1 (green lines) and  2 (black lines). 
   }
  \label{Fig2}
\end{figure*}

The external magnetic field ${\bm B}$ splits the triplet exciton sublevels and can mix them with the singlet. The resulting eigen states can be decomposed into the set of basic states (\ref{Phi})  as follows
\begin{eqnarray}\label{psi66}
\Psi_m^{(6,6)} = \sum_{j= x,y,z} C_j^{(m)}  \Psi_j^{(1s)} + C^{(m)}_{0,0} \Psi_{0,0}^{(1s)}  =
\Phi (\bm r_e,\bm r_h) \left(\sum_{j= x,y,z} C_j^{(m)}  \psi_j^{(6,6)} + C^{(m)}_{0,0} \psi_{0,0}^{(6,6)} \right)  \, . \nonumber
\end{eqnarray}
In general, the coefficients $C_j^{(m)}, C_{0,0}^{(m)}$ entering Eq.~\eqref{psi66} can be functions of the external magnetic field ${\bm B}$. 

Bounding ourselves to the cubic phase perovskites, we can choose the Cartesian coordinate frame with the $z$ axis parallel to ${\bm B}$ and write the Zeeman operator as
$$
H_Z= \mu_{\rm B} B (g_e \sigma^e_z + g_h \sigma^h_z ) \,.
$$
In this orientation of the coordinate axes, the operator $H_Z$ term can mix the exciton states only with the same projection $j_z$ on the magnetic field. Moreover, the ground-exciton spin sublevels and the wave function coefficients take the form 
\begin{eqnarray} \label{EC1} 
&&E_{1,\pm 1} = E_1 \pm  (g_e+g_h) \mu_{\rm B} B/2 \, , \quad  C_x^{1,\pm 1} = \mp \frac{1}{\sqrt{2}} \, , C_y^{1,\pm 1} = -\frac{\rm i}{\sqrt{2}} \, , C_z^{1,\pm 1} = 0 \, , \\  \label{EC0}
&& \hspace{2 cm}E_{0}^{\pm} = \frac{1}{2} \left( E_1+E_0 \pm \sqrt{\Delta_{10}^2 + [(g_e-g_h) \mu_{\rm B} B ]^2 } \right),  \,  \\ 
&& C_x^{0,\pm} = C_y^{0,\pm} =0, \quad |C_z^{0, \pm }(B)|^2 =   \frac{1}{2} \left(1 \pm  \frac{1}{\sqrt{1 +[(g_e-g_h)\mu_{\rm B}B]^2/\Delta_{10}^2}}  \right) \, , \nonumber
\end{eqnarray}
where $E_1$, $E_0$ are the excitation energies of the triplet and singlet excitons. In the weak field limit $\mu_{\rm B}B | g_e-g_h| \ll \Delta_{10}$, the mixing can be neglected and the $E_{0}^{-}$ state remains dark.  The exciton spin structure in the weak magnetic field is shown in Fig. \ref{Fig1} (b). The increasing field ${\bm B}$ modifies the energies $E_0^{\pm}$, the modification depends on the value of $g_e-g_h$ as shown in Fig. \ref{Fig2}(a,b,c). In this case, the coefficient $C^{0,\pm}_z(B)$ also depends on the magnetic field and the lowest  exciton state ${0}^{-}$ becomes optically active as shown in Fig. \ref{Fig2}(d). The exciton level structure shown in \ref{Fig2}(a,b) is similar to that considered in Ref. \cite{Belykch2022} for the weakly coupled  electron--hole pair in CsPb(Cl,Br)$_3$ perovskite nanocrystals where very slow (submilisecond) spin relaxation was observed between ${0}^{\pm}$ states.

The short-range exchange interaction Eq. \eqref{hexch} mixes also the electron-hole pairs formed by the electrons from ${\cal R}^-_6$ and  ${\cal R}^-_8$ conduction bands . The ${\cal R}^-_8 \times {\cal R}^+_6$ electron-hole pairs basis state that are  optically active in the $x,y$ or $z$ polarizations ($T_{1u}$ representation) are given by
\begin{equation}\label{8xyz}
\psi^{(8,6)}_x = \frac{1}{\sqrt{2}} ( \psi^{(8,6)}_{1,-1} -  \psi^{(8,6)}_{1,1}) \:,\:
\psi^{(8,6)}_y = \frac{\rm i}{\sqrt{2}} ( \psi^{(8,6)}_{1,-1} +  \psi^{(8,6)}_{1,1}) \:,\: \psi^{(8,6)}_z =\psi^{(8,6)}_{1,0} \:,
\end{equation}
and
\begin{eqnarray}
&&\psi^{(8,6)}_{1,1} = \frac12  \Bigl( \psi^{(e)}_{{\cal R}_8^-, 1/2} \psi^{(h)}_{{\cal R}_6^+, 1/2} - \sqrt{3} \psi^{(e)}_{{\cal R}_8^-, 3/2} \psi^{(h)}_{{\cal R}_6^+, -1/2} \Bigr) \, ,  \nonumber \\ &&\psi^{(8,6)}_{1,-1} = \frac12 \Bigl( \sqrt{3} \psi^{(e)}_{{\cal R}_8^-, -3/2} \psi^{(h)}_{{\cal R}_6^+, 1/2} -  \psi^{(e)}_{{\cal R}_8^-, -1/2} \psi^{(h)}_{{\cal R}_6^+, -1/2} \Bigr)  \nonumber \:, \\&&\psi^{(8,6)}_{1,0} = \frac{1}{\sqrt{2}} \Bigl( \psi^{(e)}_{{\cal R}_8^-, -1/2} \psi^{(h)}_{{\cal R}_6^+, 1/2} - \psi^{(e)}_{{\cal R}_8^-, 1/2} \psi^{(h)}_{{\cal R}_6^+, -1/2} \Bigr) \, .  \nonumber
\end{eqnarray}
In the basis of the $\psi^{(6,6)}_{j'=x,y,z}$ (Eq. \eqref{6xyz}) and  $\psi^{(8,6)}_{j=x,y,z}$ (Eq. \eqref{8xyz}) eigen states the matrix of the  spin part of the exchange perturbation $\hat h_{\rm exch}$ is diagonal:
\begin{equation} \label{exchjj'}
\langle \psi^{(8,6)}_j | \hat h_{\rm exch}|  \psi^{(6,6)}_{j'} \rangle = - \frac{4 \sqrt{2}}{3}\hspace{0.5 mm}  w \hspace{0.5 mm} \delta_{jj'}  = - \frac{ \sqrt{2}}{3}\hspace{0.5 mm}  \frac{\Delta_{10}} {\phi^2(0)} \hspace{0.5 mm} \delta_{jj'}  \:.
\end{equation}
In the following section we consider the mixing of the  exciton states  $\psi^{(6,6)}_{j'=x,y,z}$, Eq. \eqref{6xyz}, and  $\psi^{(8,6)}_{j=x,y,z}$, Eq. \eqref{8xyz}, caused by the interaction with acoustic phonons.

\section{EXCITON INTERACTION WITH ACOUSTIC PHONONS} \label{III}

The interaction of exciton with acoustic phonons can be described using  the deformation potential approximation providing the deformation-induced interband mixing of the electron states. The deformation tensor of the second rank, $\hat u(\rm r)$, induced by the acoustic phonon ${\bm q},\lambda$ can be written as:
\begin{equation} \label{defph}
	u_{ij}^{{\bm q},\lambda}(\bm r) =  -\frac{i}{2}\sqrt{\frac{\hbar}{2\rho \Omega_{{\bm q},\lambda}}}e^{-i{\bm q}{\bm r}}
	\left(e_i^{{\bm q},\lambda *}q_j+ e_j^{{\bm q},\lambda *}q_i\right) c^{\dag}_{{\bm q},\lambda} + c.c \, .
\end{equation}
Here $\rho$ is the mass density of the material, $V$ is the normalization volume,  ${\bm q}$ is the phonon wave vector, in the isotropic approximation $\Omega_{{\bm q},\lambda}=\Omega_{q,\lambda} = s_\lambda q$  is the phonon frequency, $s_\lambda$ is the sound velocity, $\lambda  = l, t1,t2$ denotes longitudinal and two transverse phonon branches with the polarization vectors ${\bm e}^{{\bm q},\lambda}$ given by \begin{eqnarray} \label{lt12}
	{\bm e}^{{\bm q}, l}&=&q^{-1} [q_x,q_y,q_z] \, , \: {\bm e}^{{\bm q}, t1}= (q_x^2+q_y^2)^{-1/2} [q_y,-q_x] \, ,  \\
	{\bm e}^{{\bm q}, t2}&=&q^{-1}(q_x^2+q_y^2)^{-1/2} [q_xq_z,q_yq_z, -(q_x^2+q_y^2)] \, . \nonumber
\end{eqnarray}
The $ c^{\dag}_{{\bm q},\lambda}$ and $ c_{{\bm q},\lambda}$ are the phonon creation and annihilation operators, the first term in Eq. \eqref{defph} describes the deformation tensor due to the phonon emission while the complex conjugate term is caused by the phonon absorption.

The ${\cal R}^-_8;{\cal R}^-_6$ interband part of the Bir--Pikus Hamiltonian \cite{Ivchenko,OO_book,winkler_book} including the effect of the deformation tensor $u_{ij}$ ($i,j=x,y,z$) has the form:
    \begin{equation} \label{interband}
H^{({\cal R}^-_8, {\cal R}^-_6)}(\hat{u}) = \frac{1}{\sqrt{2}}\left[ \begin{array}{cc}  d (u_{xz} - {\rm i} u_{yz}) &  \sqrt{3} b (u_{xx} - u_{yy})  - 2 {\rm i} d u_{xy} \\  - b (2u_{zz} - u_{xx} - u_{yy}) & - \sqrt{3} d  (u_{xz} - {\rm i} u_{yz})  \\ 		- \sqrt{3} d  (u_{xz} + {\rm i} u_{yz})  &  b (2 u_{zz} - u_{xx} - u_{yy}) \\ 		-\sqrt{3} b  (u_{xx} - u_{yy}) - 2 {\rm i} d  u_{xy}&   d  (u_{xz} + {\rm i} u_{yz}) 	\end{array} \right] \:,
\end{equation}
Here $b$ and $d$ are the constants of the deformation potential, $u_1=u_{xx} - u_{yy}$ and $u_2=2u_{zz}-u_{xx}-u_{yy}$. Isotropic case is described by $d=\sqrt{3} b$.  The matrix elements of the $H^{({\cal R}^-_8, {\cal R}^-_6)}(\hat{u})$ taken between  the two particle  basis functions $\psi^{(8,6)}_j$  (Eq. \eqref{8xyz}) and $\psi^{(6,6)}_{j'}$ (Eq. \eqref{6xyz}) can be written as
\begin{equation} \label{HDP2}
H_{jj'}^u = \langle \psi^{(8,6)}_j | H^{({\cal R}^-_8, {\cal R}^-_6)}(\hat{u})|  \psi^{(6,6)}_{j'} \rangle \, ,
\end{equation}
\begin{equation} \label{Hu2}
\hat{H}^u = \frac{1}{\sqrt{2}} \left[ \begin{array}{ccc} b \left( 2 u_{xx}- u_{yy} - u_{zz}\right) & \sqrt{3} d u_{xy} & \sqrt{3} d u_{zx}\\ \sqrt{3} d u_{xy}& b \left( 2 u_{yy}- u_{zz} - u_{xx}\right) & \sqrt{3} d u_{yz}  \\ \sqrt{3} d u_{zx} & \sqrt{3} d u_{yz}  & b \left( 2 u_{zz}- u_{xx} - u_{yy}\right) \end{array} \right]\:.
\nonumber
\end{equation}

\section{ Raman scattering with the exciton spin-flip assisted by phonons} \label{IV}

\subsection{Stokes and anti-Stokes processes} \label{IVA}

For simplicity, we will assume that the incident light with frequency $\omega_0$ and polarization vector ${\bm e}^0$ propagates along the normal to the surface of the substrate in the positive direction of the $z_{l}$ axis of the laboratory frame, and the scattered light with  frequency $\omega$ and polarization vector ${\bm e}$ is collected along or backward along this axis. The magnetic field is directed along $z$ axis with the angle $\theta$ to the $z_{l}$ (see Fig. 1(b)).

Then the optical matrix elements for the resonant excitation and recombination of the  exciton described by the wave functions $\Psi_m^{(6,6)}$ have the form (within multipliers)
\begin{equation}\label{opt}
M^{({\rm abs})}_n  {\cal E}_0 =  M_0 \ {\bm C}^{(n)*}\cdot {\bm e}^0 {\cal E}^0\:,\: M^{({\rm em})}_m = M^*_0\ {\bm C}^{(m)}\cdot {\bm e}^*\:.
\end{equation}
Here ${\bm C}^{(m)}$ is a vector with the components $C_x^{(m)}, C_y^{(m)}, C_z^{(m)}$, ${\cal E}_0$ is the electric-field amplitude of  the incident light, and
\[
M_0 \propto \sqrt{\frac23} d_{\rm cv} I_{\Phi}, \quad   I_{\Phi}=\int \Phi({\bm r},{\bm r}) d {\bm r} = \phi(0)\int F({\bm r}) d {\bm r}  \:,
\]
where $d_{\rm cv}$ is the interband matrix element of the dipole momentum operator $e \langle X| x | S \rangle = e \langle Y | y | S \rangle = {e \langle Z| z | S \rangle}$.

We consider the Stokes processes with the phonon initial occupation number $\bar{N}_{{\bm q},\lambda}$ and the final state with $\bar{N}_{{\bm q},\lambda}+1$ phonons and the secondary photon energy $\hbar \omega = \hbar \omega_0-\hbar \Omega_{{\bm q},\lambda} $.  For the anti-Stokes process, $\hbar \omega = \hbar \omega_0+\hbar \Omega_{{\bm q},\lambda} $ and  final phonon occupation number $\bar{N}_{{\bm q},\lambda}-1$. In equilibrium, the average value of $\bar{N}_{{\bm q},\lambda}$ is given by the Bose--Einstein function $\bar{N}_{{\bm q},\lambda} = 1/[\exp (\hbar \Omega_{{\bm q},\lambda}/k_{\rm B}T) -1]$. Then the intensity of light undergoing the exciton spin-flip scattering assisted by the  emission or absorption of the acoustic phonon has the form
\begin{eqnarray}  \label{intensity10}
	&&\hspace{- 2 mm} I^{({\rm ph})}_{+} \propto \sum_{{\bm q},\lambda} (\bar{N}_{{\bm q},\lambda}+1)|V_{f,i}^{\rm exc+ph}({\bm q},\lambda) |^2 \delta (\hbar \omega - \hbar \omega_0 + \hbar \Omega_{{\bm q},\lambda} )  ,\\
	&&\hspace{- 2 mm} I^{({\rm ph})}_{-}  \propto \sum_{{\bm q},\lambda} \bar{N}_{{\bm q},\lambda} |V_{f,i}^{\rm exc-ph}({\bm q},\lambda) |^2 \delta( \hbar \omega - \hbar \omega_0 - \hbar \Omega_{{\bm q},\lambda} ) \:, \nonumber
\end{eqnarray}

The matrix element for the double-quantum exciton spin-flip  assisted by the acoustic phonon reads
\begin{eqnarray} \label{MEexcitonS}
V_{f,i}^{\rm exc+ph}({\bm q},\lambda) =  {\cal E}_0 \sum\limits_{n m} \frac{ M^{({\rm em})}_{m}({\bm e}) V_{mn}^{{\bm q},\lambda} M^{({\rm abs})}_{n}({\bm e}^0) }{ ( E_{n} - \hbar \omega_0  - {\rm i} \hbar \Gamma_{n})( E_{m} - \hbar \omega_0+ \hbar \Omega_{{\bm q},\lambda} - {\rm i} \hbar \Gamma_{m})}  \:
\end{eqnarray}
for Stokes and
\begin{equation} \label{MEexcitonAS}
	V_{f,i}^{\rm exc-ph}({\bm q},\lambda) = {\cal E}_0  \sum\limits_{n m} \frac{ M^{({\rm em})}_{m}({\bm e}) V_{mn}^{{\bm q},\lambda} M^{({\rm abs})}_{n}({\bm e}^0) }{ ( E_{n} - \hbar \omega_0  - {\rm i} \hbar \Gamma_{n})( E_{m} - \hbar \omega_0- \hbar \Omega_{{\bm q},\lambda} - {\rm i} \hbar \Gamma_{m})}  \:.
\end{equation}
for anti-Stokes processes.
Here $n$ and $m$ denote intermediate exciton states with energies $E_n$ and $E_m$, respectively. The damping constants $\Gamma_n$ and $\Gamma_m$ for these exciton states comprise both the radiative and non radiative channels for the exciton decay, including the energy relaxation and spin-flip processes.  The matrix elements $V_{mn}^{{\bm q},\lambda}$ for the exciton spin-flip from $n$ to $m$ state with the emission  or absorption of the acoustic phonon will be calculated in Sec. \ref{defpot}.

\subsection{Resonant Raman scattering vs Resonant photoluminescence}  \label{IVB}

When a sample is illuminated by a laser, photons promote the electron subsystem to excited states: electron-hole pairs, excitons, higher-subband states of the same band etc. The initially excited states can relax to other excited states which then emit photons with frequencies different from that of the initial electromagnetic wave. In general, there are two clearly distinct phenomena, photoluminescence and light scattering. In conventional photoluminescence, the emission of photons is usually preceded by multiple transitions of the system between different real excited states, and the luminescence band of the material does not change even if one uses a different excitation wavelength. In the well-defined light scattering, the excitation frequency lies in the transparency region, excited states of the system are virtual, and the emission spectrum is tied to the initial light frequency $\omega_0$ and varies with varying this frequency. In most cases, the distinction between scattering and photoluminescence (or fluorescence) is justified. However, under certain conditions (e.g., resonant fluorescence of atoms or exciton emission in a semiconductor under resonant pumping) to distinguish between these two phenomena becomes meaningless: the light emission under resonant excitation may be interpreted not only as resonant scattering but also as resonant (or hot) photoluminescence. Bearing this in mind they use sometimes the general term `resonant secondary emission' \cite{Hizhyakov,Toyozawa,Rebane,Balkanski,Ivchenko}.

The compound matrix elements (\ref{MEexcitonS}) and (\ref{MEexcitonAS}) are derived in the third-order perturbation theory and contain a product of elementary matrix elements in the numerator and the energy denominators, which is a commonplace for describing scattering processes. Substituting Eq.~(\ref{MEexcitonS}) into Eq. \eqref{intensity10} we can rewrite the Stokes scattering intensity $I^{({\rm ph})}_+$ as
\begin{eqnarray} \label{Imm'}
&& I_+^{\rm (ph)} \propto \sum_{{\bm q}, \lambda} \delta (\hbar \omega - \hbar \omega_0 + \hbar \Omega_{{\bm q},\lambda})  \sum\limits_{m' m} M^{({\rm em})}_m ({\bm e}) M^{({\rm em})*}_{m'} ({\bm e}) \rho^{(2)}_{mm'}\:,\\ && \label{rho2} \rho^{(2)}_{mm'} = \frac{\pi}{\hbar}  \sum\limits_{n'n} (\bar{N}_{{\bm q},\lambda}+1) V^{{\bm q},\lambda}_{mn} V^{{\bm q},\lambda *}_{m'n'} \rho^{(1)}_{nn'}  \\ && \times \left[ \frac{E_m -  \hbar \omega_0+ \hbar \Omega_{{\bm q},\lambda} + {\rm i} \hbar \Gamma_m}{E_{m'} - \hbar \omega_0+ \hbar \Omega_{{\bm q},\lambda} + {\rm i} \hbar \Gamma_{m'}} \tau_m  \Delta (E_m - \hbar \omega_0 + \hbar \Omega_{{\bm q},\lambda}) \right.  \nonumber \\ && \left. ~~~+\ \frac{E_{m'} -  \hbar \omega_0 + \hbar \Omega_{{\bm q},\lambda} - {\rm i} \hbar \Gamma_{m'}}{E_{m} - \hbar \omega_0+ \hbar \Omega_{{\bm q},\lambda} - {\rm i} \hbar \Gamma_m} \tau_{m'}  \Delta (E_{m'} - \hbar \omega_0 + \hbar \Omega_{{\bm q},\lambda}) \right]\:, \nonumber  \\
&& \rho^{(1)}_{nn'} =  \frac{\pi}{\hbar}  M^{({\rm abs})}_{n}({\bm e}^0) M^{({\rm abs})*}_{n'}({\bm e}^0) {\cal E}_0^2 \label{rho1} \\ && \times \left[ \frac{E_n -  \hbar \omega_0+ \hbar \Omega_{{\bm q},\lambda} + {\rm i} \hbar \Gamma_n}{E_{n'} - \hbar \omega_0+ \hbar \Omega_{{\bm q},\lambda} + {\rm i} \hbar \Gamma_{n'}} \tau_n  \Delta (E_n - \hbar \omega_0 + \hbar \Omega_{{\bm q},\lambda}) \right.  \nonumber  \\ && \left. ~~~+\ \frac{E_{n'} -  \hbar \omega_0 + \hbar \Omega_{{\bm q},\lambda} - {\rm i} \hbar \Gamma_{n'}}{E_{n} - \hbar \omega_0+ \hbar \Omega_{{\bm q},\lambda} - {\rm i} \hbar \Gamma_n} \tau_{n'}  \Delta (E_{n'} - \hbar \omega_0 + \hbar \Omega_{{\bm q},\lambda}) \right]\:.  \nonumber
\end{eqnarray}
Here we use the notation
\begin{equation} \label{Gamma}
\tau_m = \frac{1}{2 \Gamma_m}\;,\: \Delta (E_m - \hbar \omega_0 + \hbar \Omega_{{\bm q},\lambda}) = \frac{1}{\pi} \frac{\hbar \Gamma_m}{(E_m - \hbar \omega_0 + \hbar \Omega_{{\bm q},\lambda})^2 + (\hbar \Gamma_m)^2}~~{\rm etc.}
\end{equation}
$\rho^{(1)}_{nn'}$ is the spin-density matrix of the exciton states excited by the incident light, and $\rho^{(2)}_{mm'}$ is the exciton spin-density matrix after the emission of an acoustic phonon. Since we neglect decay of the acoustic phonons, Eq.~(\ref{Imm'}) contains a Dirac $\delta$-function to account for the energy conservation requirement $\hbar \omega = \hbar \omega_0 - \hbar \Omega_{{\bm q},\lambda}$. On the contrary, Eqs. (\ref{rho2}) and (\ref{rho1}) contain smoothed delta-functions, which reduce to the exact delta-functions in the limit of $\Gamma_m, \Gamma_n \to +0$.

For the large splitting of the exciton sublevels,
\begin{equation} \label{largesplit}
|E_m - E_{m'}| \gg \hbar \Gamma_m, \hbar \Gamma_{m'}~~(m \neq m') \:,
\end{equation}
Eqs.~(\ref{Imm'})-(\ref{rho1}) reduce to
\begin{subequations}
\begin{align}
& I_+^{\rm ph} \propto \sum_{{\bm q}, \lambda} \delta (\hbar \omega - \hbar \omega_0 + \hbar \Omega_{{\bm q},\lambda})  \sum_{m} | M^{({\rm em})}_m ({\bm e})|^2 f^{(2)}_{m}\:,\\ & f^{(2)}_{m} = \frac{2 \pi}{\hbar}  \sum_{n} (\bar{N}_{{\bm q},\lambda}+1) |V^{\lambda}_{mn}|^2 \tau_{m}  f^{(1)}_n  \Delta (E_m - \hbar \omega_0 + \hbar \Omega_{{\bm q},\lambda})\:,  \\ & f^{(1)}_n =  \frac{2 \pi}{\hbar} |M^{({\rm abs})}_{n}({\bm e}^0) |^2 {\cal E}^2_0 \tau_{n} \Delta( E_n - \hbar \omega_0)\:.
\end{align}
\end{subequations}
The conditions (\ref{largesplit})  mean that the exciton sublevels can be considered as isolated from each other with the probability of occupation $f^{(1)}_n$ and $f^{(2)}_m$, respectively, and the coherence between them can be neglected. Therefore, in this particular case, the resonant Raman scattering can be equivalently described as a resonant excitation of the $E_n$ states, phonon-assisted energy relaxation to the $E_m$ states and the subsequent radiative recombination.

If the conditions (\ref{largesplit}) are invalid, the coherence between the exciton states can be remarkable and the density-matrix formalism gets important. Equations (\ref{Imm'})-(\ref{rho1}) may be considered as an extension of the theory of elastic resonant secondary emission \cite{JETP1977} to treat spin-dependent inelastic resonant processes.
\subsection{The acoustic-phonon mediated  exciton spin-flip  } \label{defpot}
The acoustic-phonon assisted exciton spin-flip can be calculated in the second-order perturbation theory taking into account electron-hole exchange interaction $H_{\rm exch}$, Sec.~\ref{setup}, and the electron-phonon deformation potential $H_{\rm DP}$, Sec.~\ref{III}. For the ${\cal R}^-_8 \times {\cal R}^+_6$ electron-hole pairs one can neglect the Coulomb interaction as it is much smaller than the spin-orbit splitting energy $\Delta$. The matrix element of the spin-flip transition can be written as
\begin{equation} \label{Vjj'}
V^{{\bm q},\lambda}_{mn} = \sum_{jj' = x,y,z}C^{(m)*}_j {\cal V}^{{\bm q},\lambda}_{jj'} C^{(n)}_{j'} \:,
\end{equation}
where the coefficients $C^{(m)}, C^{(n)}$ are given by Eqs.~(\ref{EC1}) and (\ref{EC0}), and $V^{{\bm q},\lambda}_{jj'}$ is the compound matrix of the two-quantum transition
\begin{equation} \label{Vq}
{\cal V}^{{\bm q},\lambda}_{jj'} = \frac{1}{\Delta} \sum_{k} \Bigl( \langle \Psi_j^{(6,6)} | H_{\rm exch}| k \rangle \langle k | H_{\rm DP}|  \Psi_{j'}^{(6,6)} \rangle + \langle \Psi_{j}^{(6,6)} | H_{\rm DP}| k \rangle \langle k | H_{\rm exch}|  \Psi_{j'}^{(6,6)} \rangle \Bigr)\:.
\end{equation}
Here the index $k$ stands for the two particle excitations involving an electron in the higher conduction band ${\cal R}_8^-$ and a hole in the valence band ${\cal R}_6^+$. In the basis (\ref{8xyz}) the full set of these excitations is described by
\[
| k \rangle = \frac{ {\rm e}^{{\rm i} ({\bm k_e} {\bm r_e}+ {\bm k_h}{\bm r_h})}}{\sqrt{V}}\ {\bm C}^{(8,j)} ~~(j = x,y,z)\:,
\]
where ${\bm k_e},{\bm k_h}$ are the electron and hole wave vectors, and ${\bm C}^{(8,j)}$ is the vector with the components $C^{(8,j)}_{j'} = \delta_{jj'}$. Note, that the kinetic energies of the  ${\cal R}_8^-$  electrons and  ${\cal R}_6^+$ holes corresponding to the finite values of ${\bm k_e},{\bm k_h}$ are also neglected in the denominator $\Delta$ in Eq. \eqref{Vq}. Taking into account the identity
\[
\sum_{k} | k \rangle \langle k |  = 1\:,
\]
we obtain for the first sum in Eq.~(\ref{Vq})
\begin{equation} \label{sumk}
\sum_{k}  \langle \Psi_j^{(6,6)} |H_{\rm exch}| j, k \rangle \langle j,k | H_{DP}|  \Psi_{j'}^{(6,6)} \rangle = \int d{\bm r}  \left[ \Phi^2 ({\bm r},{\bm r}) \langle
 \psi_j^{(6,6)} | \hat h_{\rm exch}|  \psi_{j}^{(8,6)} \rangle H^u_{jj'}({\bm r}) \right]  \, .
\end{equation}
The second sum in Eq.~(\ref{Vq}) is treated in a similar way.

Using the method of invariants based on symmetry considerations \cite{OO_book} we can represent the right-hand side of Eq.~(\ref{Vq}) in the form
\begin{equation} \label{Xi}
{\cal V}^{{\bm q},\lambda}_{jj'} = \frac{\Delta_{10} }{\Delta} {\cal F}_{\bm q} \left[ \left( D_1 \sum_i U^{{\bm q},\lambda }_{ii} + D_2 U^{{\bm q},\lambda }_{jj'} \right) \delta_{jj'} + D_3 (1 - \delta_{jj'}) U^{{\bm q} \lambda }_{jj'} \right]\:.
\end{equation}
Here the amplitude of the acoustic phonon deformation tensor of Eq.~\eqref{defph} is given for the Stokes process by
\begin{equation} \label{Uq}
U^{{\bm q}, \lambda }_{jj'} =  -\frac{\rm i}{2}\sqrt{\frac{\hbar q }{2\rho s_{\lambda}}} Q^{ \hat{\bm q}, \lambda }_{jj'}\, , \quad
Q^{\hat{\bm q}, \lambda }_{jj'} =\left( \hat{q}_j e_{j'}^{{\bm q},\lambda*} + \hat{q}_{j'} e_j^{{\bm q},\lambda* }\right)\:,
\end{equation}
$\hat{\bm q} = {\bm q}/q$, and
\begin{equation}
{\cal F}_{\bm q} = \int F^2({\bm r}) {\rm e}^{-{\rm i} {\bm q}{\bm r}} d{\bm r}\: \nonumber
\end{equation}
is the electron-phonon scattering form-factor describing the dependence of the scattering amplitude on the spatial spread of the exciton center of mass, the function $ F({\bm r})$ is defined in Eq.~(\ref{fact}).  In deriving Eqs.~(\ref{Xi}), (\ref{Uq}), use was made of Eq. (\ref{exchjj'}) for the matrix elements of $\hat h_{\rm exch}$.

The constants   $D_l$ ($l=1,2,3$) are deduced from Eqs.~(\ref{Vq}) and (\ref{sumk}) and read
\begin{equation} \label{D2D3}
D_2 = - 2 b, D_3 = -  \frac{2 d}{\sqrt{3}}, D_1 = - \frac{D_2}{3} = \frac{2 b}{3}\:.
\end{equation}
For the anti-Stokes process with the phonon absorption one has to use ${\cal V}^{{\bm q},\lambda*}_{jj'}$ in Eq.~(\ref{Vjj'}).

\section{Analysis of the spin-flip scattering with $\Delta j_z = 2, 1, 0$} \label{DC}

Equations (\ref{Imm'})-(\ref{rho1}),  (\ref{Vjj'}),  (\ref{Vq}), and  (\ref{Uq}) form a set that allows one to calculate the frequency and polarization dependencies of the efficiency of the resonant Raman scattering with the flip of exciton angular momentum component by $\Delta j_z =j_z^n-j_z^m= 2, 1, 0$, where $n$ and $m$ denote the first and second intermediate exciton states.

Because of the exciton damping the spread $\Delta q$ of wave vectors of acoustic phonons involved in the scattering is $\sim \Gamma_m/s_{\lambda}$.  The exciton damping rate $\Gamma_m$ is assumed to be small, such that the scattering form-factor (\ref{form}) changes negligibly within the interval $ \Delta q$. In this case the $\Delta$-function in Eq.~(\ref{Gamma}) can be replaced by the exact $\delta$-function, and the phonon energy $\hbar \Omega_{q,\lambda}$ is fixed to $E_n - E_m = \hbar (\omega_0 - \omega)$. The values of $q$ are different for the longitudinal and transverse phonons with the same energy $\hbar \Omega = \hbar (\omega_0 - \omega) $ because of the different sound velocity $s_l$ and $s_t$. For simplicity of estimates, hereafter we neglect  this difference and assume $s_l=s_t=s$ as well as we neglect the cubic anisotropy of the phonon dispersion. Moreover, we assume the $F({\bm r})$ function describing the center of mass exciton localization to be spherically symmetric and replace ${\cal F}_{\bm q}$ by ${\cal F}_q$:
\begin{equation}
\label{form}
{\cal F}_q = \int F^2({r}) {\rm e}^{-{\rm i} {\bm q}{\bm r}} d{\bm r} = 
4 \pi \int_0^\infty F^2({r}) j_0 (q r) r^2d{r} \, , 
\end{equation}
where $r=|{\bm r} - {\bm r}_0|$ is the distance from the exciton center of mass to the localization center, and $j_0(x) = \sin x/x$ is the spherical Bessel function. 

For definiteness, we estimate here the Stokes scattering intensity $I^{({\rm ph})}_+$ in which case the temperature dependence is characterized by a factor of
\[
\bar{N}_{{\bm q}, \lambda} + 1 = \left[ 1 - \exp{\left(- \hbar (\omega_0 - \omega)/k_{\rm B} T\right)}\right]^{-1} \:.
\]

{\it SFRS with} $\Delta j_z=2$. For the spin-flip transition with $\Delta j_z = 2$, see arrows $X_{\gamma}$ in Fig.~2, the Raman shift is a linear function of B: $\hbar (\omega_0 - \omega) = E_n - E_m = g \mu_{\rm B} B \equiv \Delta_{\gamma}$.  Here $E_n=E_{1,+1} $ and $E_m=E_{1,-1}$ are the photoexcited exciton states before and after the phonon emission. Under the resonant excitation of the $E_{1,+1}$ state, a difference between the concepts ``scattering'' and ``luminescence'' is lost. The participation of acoustic phonons in the scattering process under consideration does not mean that this process can be interpreted as Brillouin scattering. In the conventional Brillouin scattering the light wave is diffracted by the dynamic grating induced by an acoustic wave. As a result, the frequency shift  is given by $\omega_0 - \omega = s_{\lambda}|{\bm k}- {\bm k}_0|$, where ${\bm k}_0, {\bm k}$ are the light wave vectors of the initial and scattered light. In contrast, in the light scattering with the exciton spin-flip and emission of an acoustic phonon, the frequency shift $\Delta_{\gamma}/\hbar$ is unrelated to the photon wave vectors.

 Using Eqs. (\ref{EC1}), (\ref{Vjj'}), (\ref{Xi}) and (\ref{D2D3}), we obtain the matrix element $V_{-1,+1}^{{\bm q},\lambda}$ for the $X_\gamma$ transition 
  \begin{eqnarray}
    V_{-1,+1}^{{\bm q},\lambda} = -\frac{1}{2}\left[{\cal V}^{{\bm q},\lambda}_{xx}- {\cal V}^{{\bm q},\lambda}_{yy} +i ({\cal V}^{{\bm q},\lambda}_{xy} + {\cal V}^{{\bm q},\lambda}_{yx})\right]
   =\frac{\Delta_{10} }{\Delta} {\cal F}_{q} \left[ b (U_{yy}^{{\bm q},\lambda} - U_{xx}^{{\bm q},\lambda}) + {\rm i}\frac{2d}{\sqrt{3}} U_{xy}^{{\bm q},\lambda} \right] \, . 
 \end{eqnarray}
To find the scattering efficiency as a function of the system parameters we average the squared modulus of $V_{-1,+1}^{{\bm q},\lambda}$ over the directions $\hat{\bm q}$ of the phonon propagation, sum over the phonon modes and use the identity valid for $s_l=s_t=s$:
$$\sum_{\hat{\bm q},\lambda} Q_{ij}^{\hat{{\bm q}},\lambda}Q_{i'j'}^{\hat{{\bm q}},\lambda} =  \frac{1}{2}\left(  \delta_{jj'} \delta_{i i'} +  \delta_{j i'} \delta_{i j'}  \right)\:.$$ 
The result reads 
$$\sum_{\hat{\bm q},\lambda}   |V_{-1,+1}^{{\bm q},\lambda}|^2  = 
\left(\frac{ \Delta_{10}}{\Delta }\right)^2 {\cal F}_q^2 \frac{\hbar q}{4\rho s}  D_\gamma^2 \, , \quad  D_\gamma^2= b^2+\frac{d^2}{3} \, .
$$
The integration over the absolute values ${\bm q}$ adds a factor $q^2 = (\Delta_\gamma/\hbar s)^2$, and 
the  intensity of the $X_{\gamma}$ scattering can be finally written as
\begin{eqnarray}  \label{intensityX2}
I_{+}^{(\gamma)} \propto \rho(\hbar \omega_0) (N_\gamma+1) \delta(\hbar \omega - \hbar \omega_0 +\Delta_\gamma) M_0^4 {\cal E}_0^2 \left(\frac{ \Delta_{10}}{\Delta }\right)^2 \frac{\Delta_\gamma^3 }{ \rho s^5} \frac{\tau_{1,+1}\tau_{1,-1}}{\hbar^5 }  {\cal F}_{q}^2 D_\gamma^2 \,  .
\end{eqnarray}
Here $\rho(\hbar \omega_0)$ is the density of localized exciton states of energy $\hbar \omega_0$  and the phonon emission factor $N_\gamma+1 = \left[ 1 - \exp{\left(- \Delta_\gamma/k_{\rm B} T\right)}\right]^{-1}$. The form-factor ${\cal F}_{q = \Delta_\gamma/\hbar s}$ can be found from Eq. (\ref{form}) by introducing the shape of the center-of-mass localization envelope $F(r)$ of the localized exciton with the energy $\hbar \omega_0$.  
   
The dependence of the intensity on the magnetic field is governed  by the function $N_\gamma+1$, the factor $\Delta_\gamma^3 \propto B^3$ in the nominator and the $q$-dependence of  ${\cal F}_{q}^2$. The latter is determined by the product of $q a$, where $a$ is the localization radius. For example, if we take $F(r) \propto \exp(-r/a)/r$ for exciton localization at a spherical short-range potential, the function ${\cal F}_{q}$ varies as $\arctan(qa/2)/qa$.  In the limit $qa \gg 1$, ${\cal F}_{q}^2 \propto (qa)^{-2}$, and the magnetic field dependence of the intensity becomes linear instead of cubic. In addition, the exciton lifetimes $\tau_{1,\pm1}$ may also depend on the magnetic field.

While deriving an estimation for $I_{+}^{(\gamma)}$  we have concentrated on the material parameters entering Eq.~(\ref{intensityX2}) and ignored the light polarization. The dependence of $I^{(\gamma)}_+$  on the light polarization and the magnetic-field direction is described by the function
 \begin{equation} \label{PX2}
 P_{2}({\bm e},{\bm e}^0) = |{\bm C}^{1,-1}\cdot {\bm e}^*|^2|({\bm C}^{1,+1})^*\cdot {\bm e}^0|^2 = \frac14 \left( 1 - |{\bm e} {\bm b}|^2 - {\bm \kappa} {\bm b}\right) \left( 1 - |{\bm e}^0 {\bm b}|^2 + {\bm \kappa}_0 {\bm b} \right) .
 \end{equation}
Here ${\bm \kappa} = {\rm i} ({\bm e} \times {\bm e}^*) = P_{\rm circ}{\bm k}/{k}$, 
$P_{\rm circ}$ is the degree of circular polarization, ${\bm \kappa}_0$ is defined in the same way for the incident light, and ${\bm b} = {\bm B}/B$.  The selection rules (\ref{PX2}) certainly coincide with those for (a) the double scattering with simultaneous spin flips of the resident electron and hole and the change of the sum of their spin projections on the magnetic field by 2, as well as for (b) the spin-flip scattering via the biexciton intermediate state \cite{Rodina2022}. In that reference one can find a detailed analysis of the function (\ref{PX2}) on the polarization of the initial and scattered light and on the magnetic-field orientation.

{\it SFRS  with} $\Delta j_z=1$. In a weak magnetic field, this kind of spin-flip transitions marked as $X_\alpha$ and $X_\beta$ in  Figs.~\ref{Fig1} and \ref{Fig2} makes a clear distinction between the two scattering mechanisms, namely, the acoustic-phonon involved mechanism and that with simultaneous  spin-flips of resident electron and hole. The former allows Raman scattering photon energy shifts $E_{1,1} - E_{1,0}$ and $E_{1,0} - E_{1,-1}$, in weak fields the shifts coincide and are equal to $\Delta_{\alpha}= \Delta_{\beta}=\Delta_{\gamma}/2 = g\mu_{\rm B} B/2$, Fig.~\ref{Fig1}(b). The latter mechanism does not provide a shift by $g\mu_{\rm B} B/2$, only by $g\mu_{\rm B} B$ or by $|g_e-g_h|\mu_{\rm B} B$.

We analyze the selection rules in the following geometry: the normal light incidence in the positive direction of the $z_l$ axis of the laboratory frame (see Fig. \ref{Fig1}(b)) and a backward registration of scattered light, $\bar z_l$. In the tilted geometry with $\theta \ne 0,\pi$, the exciton states $+1$ and $0^{+}$ can be excited by the circularly or linearly polarized light, the corresponding matrix elements are proportional to $-e^0_x + {\rm i} e^0_y$ and ${\bm e}^0{\bm b}$, and the emission matrix elements for the $0^{+}$ and $-1$ states are proportional to ${\bm e}^*{\bm b}$ and $-(e^*_x - {\rm i} e^*_y)$, see Eqs.~(\ref{EC1}), (\ref{EC0}) and (\ref{opt}). 

If $\Delta_{\alpha}, \Delta_{\beta} \ll \hbar \Gamma_{J=1}$, then one should take into account an interference between the channels $X_{\alpha}$ and $X_{\beta}$, and the polarization selection rule reads  
\begin{eqnarray}  \label{intensityX1}
P_{1}({\bm e},{\bm e}^0) = |{\bm e}^* \times {\bm e}^0|^2 \sin^2 \theta \, .
	\end{eqnarray}
It is the same as for the spin-flip of a single resident carrier (electron or hole) or the biexciton mechanism of SFRS with  a change of the exciton spin component $j_z$ by 1 \cite{Rodina2022}. Contrary to the scattering with $\Delta j_z=2$, the process with $\Delta j_z=1$ is forbidden in the Faraday geometry ($\theta = 0, \pi$) and allowed only in a tilted magnetic field or the Voigt geometry ($\theta = \pi/2$). The scattering in the crossed linear (${\bm e} \perp {\bm e}^0$) and co-circular ($\sigma_+, \sigma_+$~or~$\sigma_-, \sigma_-$) configurations is allowed and occurs with the equal probability proportional to $\sin^2{\theta}$, while in the co-linear (${\bm e} \parallel {\bm e}^0$) and crossed circular ($\sigma_-, \sigma_+$~or~$\sigma_-, \sigma_+$) configurations the scattering is 
prohibited. 

In the opposite limiting case of the strong Zeeman splitting, $  \Delta_{\alpha}, \Delta_{\beta} \gg \hbar \Gamma_{J=1}$, the processes $X_{\alpha}$ and $X_{\beta}$ do not interfere, and the resulting intensity can be considered as a sum of the partial intensities from the these two channels with the polarization rules
\begin{eqnarray}  \label{intensityX11}
&&P_{X_{\alpha}}({\bm e},{\bm e}_0) =  |{\bm e}{\bm b}|^2 |({\bm C}^{1,+1})^*{\bm e}^0|^2 = \frac12 \left( 1 - |{\bm e}^0 {\bm b}|^2 + {\bm \kappa}_0 {\bm b} \right)|{\bm e}{\bm b}|^2 \, , \\ &&P_{X_{\beta}}({\bm e},{\bm e}_0) =  |{\bm C}^{1,-1}{\bm e}^*|^2|{\bm e}^0{\bm b} |^2 = \frac12 \left( 1 - |{\bm e} {\bm b}|^2 - {\bm \kappa} {\bm b}\right)|{\bm e}^0{\bm b} |^2 \, . \nonumber
	\end{eqnarray}
Both channels are forbidden in the pure Faraday geometry similarly to eq.~(\ref{intensityX1}). However, in the tilted magnetic field, the polarization functions $P_{X_{\alpha}}({\bm e},{\bm e}_0)$ and $P_{X_{\beta}}({\bm e},{\bm e}_0)$ taken separately and together and different from the selection rules (\ref{intensityX1}). Note that, for the strong splitting between the exciton sublevels, the selection rules for single-particle SFRS may change and allow the process in crossed circular polarizations ($\sigma_-, \sigma_+$~or~$\sigma_-, \sigma_+$) due to a fast Larmor precession of the unpaired spin in the photoexcited trion \cite{Rodina2022}. 

With the increasing magnetic field the energy shifts $\Delta_{\alpha}$ and $\Delta_{\beta}$ become different and depend nonlinearly on the magnetic field, Fig. \ref{Fig2}(b,c). In the limit of strong magnetic fields, $|g_e - g_h| \mu_{\rm B} B \gg \Delta_{10}$, the shifts tend to 
\begin{eqnarray}
&&\Delta_{\alpha} = E_{1,1} - E^+_0 \to \frac{g_e + g_h - |g_e - g_h|}{2} \mu_{\rm B} B,\nonumber\\&& \Delta_{\beta} = E^+_0 - E_{1,-1} \to  \frac{g_e + g_h + |g_e - g_h|}{2}  \mu_{\rm B} B\:.
\end{eqnarray}
Particularly, for the positive difference $g_e - g_h$, the values $\Delta_{\alpha}, \Delta_{\beta}$ tend to the single-particle splittings $g_h \mu_{\rm B} B$ and  $g_e \mu_{\rm B} B$, respectively. Moreover, the new transitions  $X'_{\alpha}$ and $X'_{\beta}$ arise involving the magnetic-field-activated $0^{-}$ exciton, as shown in Fig.~\ref{Fig2}(b,c). Their polarization rules are also described by Eqs.~(\ref{intensityX11}), and their intensities increase proportionally to $|C_z^{-}(B)|^4$, while the intensities of $X_{\alpha}$ and $X_{\beta}$ processes decrease as $|C_z^{+}(B)|^4$. If $|\mu_{\rm B}B(g_e-g_h)| \gg |\Delta_{10}|$, the energy shifts and intensities for the processes $X'_{\alpha}$ and $X'_{\beta}$ excitons coincide with those for the $X_{\alpha}$ and $X_{\beta}$ processes.

{\it SFRS with} $\Delta j_z=0$. In a stronger magnetic field one can observe the $X_{\delta}$ transition between  the $E_n=E_{0}^{+}$ and  $E_m=E_{0}^{-}$ states without a change of the angular momentum component $j_z$, Fig. 2(b,c). In this additional transition, the Raman shift  is dependent on the $g$ factor difference $g_e-g_h$ as
\begin{equation} \label{E0E0}
\Delta_{\delta} = \sqrt{\Delta^2_{10} + [(g_e - g_h)\mu_{\rm B} B]^2}\:,
\end{equation}
the intensity is proportional to 
\begin{equation} \label{CzCz}
|C_z^{+}(B)|^4 |C_z^{-}(B)|^4  = \frac{1}{16} \left\{ 1 - \frac{1}{1 + [(g_e - g_h) \mu_{\rm B} B/\Delta_{10}]^2 } \right\}^4 \:,\end{equation} 
the polarization rule reads
\begin{equation}  \label{intensityX0}
P_{X_{\delta}}({\bm e},{\bm e}_0) = |{\bm e}^0{\bm b}|^2|{\bm e}{\bm b}|^2 \, ,
	\end{equation}
and the process is forbidden in the pure Faraday geometry. The SFRS reaches maximum in the Voight geometry ${\bm B} \perp z_l$ and the colinear configuration $ {\bm e}^0 \parallel {\bm e} \parallel {\bm B}$. Apparently, if $g_e$ and $g_h$ have the same signs, values of the factor (\ref{CzCz}) and frequency shift $E^+_0 - E_0^-$ can be small which makes it difficult to observe the Raman scattering with $\Delta j_z=0$, the $X_\delta$ exciton transition in Fig.~2. However, in perovskites with $g_e$ and $g_h$ of opposite signs, Fig.~3(c), this kind of scattering may be effective in a sufficiently strong magnetic field. It is worth to mention here that the light scattering with a flip-flop of the resident electron and hole can be observed already in a weak field as discussed in \cite{Rodina2022}.

\section{Conclusion} \label{DC2}

We have studied the acoustic phonon mechanism of SFRS in perovskite semiconductors. To this end, we  have derived the matrix elements of acoustic-phonon induced spin-flip transitions between Zeeman sublevels of the localized 
 ground-state exciton  ${\cal R}^-_6 \times {\cal R}^+_6$. These transitions are treated in the second-order perturbation theory taking into account electron-hole exchange interaction and the electron-phonon interaction in the deformation potential approximation: due to one of these interactions the ground state exciton is transferred to a virtual electron-hole pair with the electron lying in the higher conduction band ${\cal R}^-_8$, and due to the other interaction, the virtual excitation is returned to the ${\cal R}^-_6 \times {\cal R}^+_6$ exciton state with a changed angular momentum projection $j_z$ on the magnetic field. We have analyzed the selection rules and efficiency of the phonon-involved Raman scattering with the change of $j_z$ by 2 or 1 as well as the scattering with a transition with $\Delta j_z = 0$, between the exciton sublevels $E_0^+$ and $E_0^-$ mixed by the magnetic field.  

The phonon mechanism considered here complements the analysis of two other mechanisms of SFRS in persovskites presented in Ref.~\cite{Rodina2022}. They are (a) light scattering with double spin reversal of isolated resident electron and hole, and (b) spin-flip of the photoinduced exciton via its photoexcitation to a biexciton state by a second photon followed by emission of a scattered (secondary) photon by the biexciton. The most important difference between the manifestation of the mechanism (a) and the phonon mechanism, let it be the mechanism (c), is respectively an absence and a presence of the Raman half-shift $g\mu_B B/ 2$. The other difference is a presence (a) and an absence (c) of the transitions with  the Raman half-shift $|g_e-g_h|\mu_B B/ 2$ at low magnetic fields. The mechanisms (c) and (b) differ in linear and nonlinear dependence of the scattering efficiency on the incident light intensity. The  intensity of the phonon-assisted scattering is expected to increase with the magnetic field, however the power of this increase depends on the radius of exciton localization. The three mechanisms (a), (b) and (c) complete the picture of resonant spin-flip Raman scattering mediated by localized excitons in perovskites. The considered mechanism of the acoustic phonon-assisted transitions between the exciton Zeeman sublevels may be also important for analysis of the spin relaxation processes for localized excitons in perovskite semiconductor nanostructures \cite{Belykch2022,Sercel2019,Yugova}.

{\bf Funding:} The work of A.V.R. on the analytical consideration of the exciton fine structure and spin-flip transitions was supported by the Russian Science Foundation (RFS) grant No. 23-12-00300, and the work of E.L.I. work on the symmetry analysis of the exciton spin-flip processes was supported by the RFS grant No. 23-12-00142.


\bibliographystyle{cas-model2-names}


\end{document}